\documentclass[%
 reprint,
 superscriptaddress,
 amsmath,amssymb,
 aps,
 prl,
]{revtex4-2}

\usepackage{graphicx}
\usepackage{floatrow}
\usepackage{color}
\usepackage{float}
\usepackage{subfig}
\usepackage{caption}
\usepackage{comment}
\usepackage{dcolumn}
\usepackage{bm}
\begin{document}

\preprint{APS/123-QED}

\title{Enhanced timing of a 113 km O-TWTFT link with digital maximum likelihood estimation process}

\author{Yu-Chen Fang}
\affiliation{Hefei National Research Center for Physical Sciences at the Microscale and School of Physical Sciences, University of Science and Technology of China, Hefei, China}
\affiliation{Shanghai Research Center for Quantum Sciences and CAS Center for Excellence in Quantum Information and Quantum Physics, University of Science and Technology of China, Shanghai, China}
\affiliation{Hefei National Laboratory, University of Science and Technology of China, Hefei, China}
\author{Jian-Yu Guan}
\affiliation{Shanghai Research Center for Quantum Sciences and CAS Center for Excellence in Quantum Information and Quantum Physics, University of Science and Technology of China, Shanghai, China}
\affiliation{Hefei National Laboratory, University of Science and Technology of China, Hefei, China}
\author{Qi Shen}
\author{Jin-Jian Han}
\affiliation{Hefei National Research Center for Physical Sciences at the Microscale and School of Physical Sciences, University of Science and Technology of China, Hefei, China}
\affiliation{Shanghai Research Center for Quantum Sciences and CAS Center for Excellence in Quantum Information and Quantum Physics, University of Science and Technology of China, Shanghai, China}
\affiliation{Hefei National Laboratory, University of Science and Technology of China, Hefei, China}
\author{Lei Hou}
\affiliation{Shanghai Research Center for Quantum Sciences and CAS Center for Excellence in Quantum Information and Quantum Physics, University of Science and Technology of China, Shanghai, China}
\affiliation{Hefei National Laboratory, University of Science and Technology of China, Hefei, China}
\author{Meng-Zhe Lian}
\affiliation{Hefei National Research Center for Physical Sciences at the Microscale and School of Physical Sciences, University of Science and Technology of China, Hefei, China}
\affiliation{Shanghai Research Center for Quantum Sciences and CAS Center for Excellence in Quantum Information and Quantum Physics, University of Science and Technology of China, Shanghai, China}
\affiliation{Hefei National Laboratory, University of Science and Technology of China, Hefei, China}
\author{Yong Wang}
\affiliation{Xinjiang Astronomical Observatory, Chinese Academy of Sciences, Urumqi 830011, China}
\author{Wei-Yue Liu}
\affiliation{Shanghai Research Center for Quantum Sciences and CAS Center for Excellence in Quantum Information and Quantum Physics, University of Science and Technology of China, Shanghai, China}
\affiliation{Hefei National Laboratory, University of Science and Technology of China, Hefei, China}
\affiliation{Faculty of Information Science and Engineering, Ningbo University, Ningbo 315211, China}
\author{Ji-Gang Ren}
\author{Cheng-Zhi Peng}
\author{Qiang Zhang}
\author{Hai-Feng Jiang}
\author{Jian-Wei Pan}
\affiliation{Hefei National Research Center for Physical Sciences at the Microscale and School of Physical Sciences, University of Science and Technology of China, Hefei, China}
\affiliation{Shanghai Research Center for Quantum Sciences and CAS Center for Excellence in Quantum Information and Quantum Physics, University of Science and Technology of China, Shanghai, China}
\affiliation{Hefei National Laboratory, University of Science and Technology of China, Hefei, China}


\begin{abstract}
Optical two-way time-frequency transfer (O-TWTFT), employing linear optical sampling and based on frequency combs, is a promising approach for future large-scale optical clock synchronization. It offers the dual benefits of high temporal resolution and an extensive unambiguous range. A critical challenge in establishing long-distance free-space optical links is enhancing detection sensitivity. Particularly at ultra-low received power levels, the error caused by time extraction algorithms for linear optical sampling becomes a significant hindrance to system sensitivity, surpassing the constraints imposed by quantum limitations. In this work, we introduce the Complex Least Squares (CLS) method to enhance both the accuracy and sensitivity of time extraction. Unlike most previous methods that relied solely on phase information, our scheme utilizes a maximum likelihood estimation technique incorporating both amplitude and phase data. Our experiments, conducted over a 113 km free-space link with an average link loss of up to 100 dB, achieved a record minimum received power of 0.1 nW, which is over ten times lower than previous benchmarks. The precision also approaches the quantum limitation.
\end{abstract}

\maketitle
 

\section{\label{sec:level1}introduction}
The advancement of optical clocks has necessitated the development of high-precision time and frequency transfer systems on a global scale. State-of-the-art optical clocks have achieved remarkable instability and uncertainty levels of $10^{-19}$~\cite{brewer2019al+, bothwell2019jila, nicholson2015systematic, huntemann2016single, aeppli2024clock}, surpassing the best primary frequency standards by over an order of magnitude. These precision levels are essential for applications such as the redefinition of the second~\cite{dimarcq2024roadmap} and precision geodesy~\cite{mehlstaubler2018atomic, lisdat2016clock}. While fiber-based links can meet these demands, their global deployment is hindered by cost and complexity in the near future.

Current global time and frequency transfer relies on satellite-based relays. But none meets the requirement of optical clock frequency transfer. Ground-satellite links require a large unambiguous timing range for continuous operation due to the noisy and unstable atmospheric conditions. Traditional microwave signals, used as carriers, limit long-term instability to a few picoseconds due to their low timing resolution. Time transfer by laser link (T2L2), which uses pulse lasers and photodetectors, faces similar limitations~\cite{exertier2010status, exertier2014time}.

The most viable solution is the optical two-way time-frequency transfer (O-TWTFT), which employs the two-way transmission of coherent pulse laser trains, i.e., optical frequency combs (OFCs), combined with linear optical sampling (LOS)~\cite{giorgetta2013optical}. The LOS method involves interfering a local OFC with a received OFC of a slightly different repetition rate, allowing the local comb's pulses to continuously traverse the signal pulse. The OFC's nanosecond-level pulse interval provides a large unambiguous timing range, facilitating link reconnection after interruptions.

Achieving a ground-satellite O-TWTFT link requires overcoming transmission losses ranging from tens to over a hundred dB, depending on the distance and optical transceiver's effective aperture. In 2022, we demonstrated an 89-dB, 113-km O-TWTFT link with an instability of $10^{-19}$ at 10,000 seconds, using an OFC power of up to 1 W and detection sensitivity in the nanowatt range~\cite{shen2022free}. This high tolerance for loss is suitable for GEO/IGSO satellite-to-ground links but falls short for deep space applications like moon-earth links.

To enhance loss tolerance, the NIST group pioneered the use of optical interference between two OFCs with the same repetition rate, employing a programmable OFC for continuous signal comb tracking~\cite{caldwell2022time}. This approach increases sensitivity to about 330 femtowatts, enabling a 300-kilometer optical link with up to 102 dB loss~\cite{caldwell2023quantum}. While this system can handle the high loss of a moon-earth link, the rapid Doppler effect in real links necessitates rapid and extensive phase tuning of the programmable OFC for continuous tracking, presenting further challenges and system complexity.

In this study, we introduce an innovative approach to bolster the system's resilience to signal loss, leveraging a software-based enhancement of the traditional LOS technology. This method not only retains the inherent benefits of LOS in handling Doppler shifts~\cite{bergeron2019femtosecond} but also significantly amplifies its sensitivity. The primary bottleneck in detection sensitivity within the LOS framework stems from the time extraction algorithm, a limitation that surpasses even the quantum noise floor. We present a novel time extraction algorithm designed to enhance both the precision and sensitivity of timing determination. This solution offers a cost-efficient upgrade to the existing LOS systems, requiring nothing more than additional data processing capabilities.

This paper is structured as follows: In Section II, we present our novel time extraction algorithm tailored for LOS results. Section III details our experimental setup and parameters. In Section IV, we compare the performance of our algorithm against previous methods. Finally, Section V provides a concise summary of our findings.

\section{Time extraction algorithm}
\begin{figure}[htb]\centering
\sidesubfloat[]{\includegraphics[width=\columnwidth]{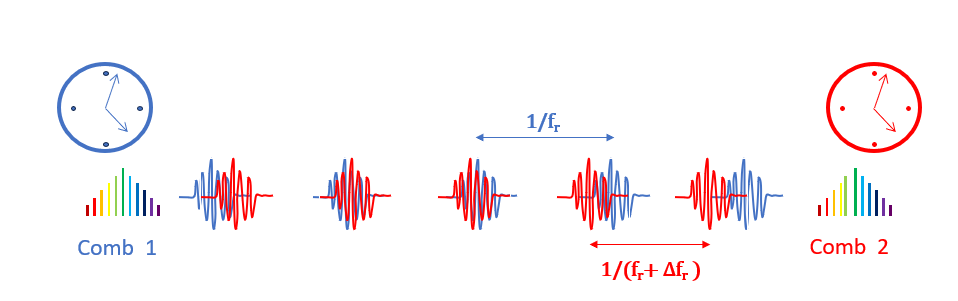}\label{fig:pic1}}
\hfill
\sidesubfloat[]{\includegraphics[width=\columnwidth]{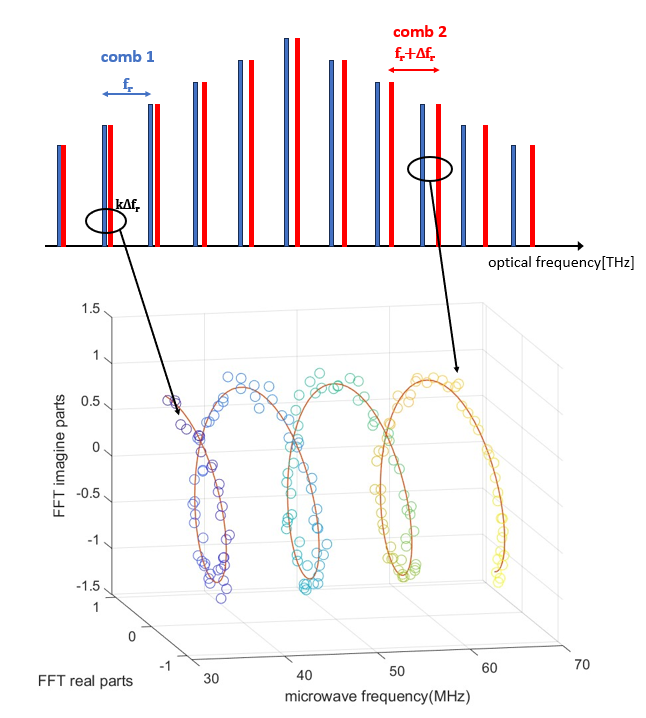}\label{fig:pic2}}
\caption{(a) The time domain picture of O-TWTFT: The two terminals exchange frequency comb pulse trains, where their repetition rates, denoted as \( f_{r1} and f_{r2} \), are offset by a small frequency difference \( \Delta f_r \). (b) The upper part of the figure presents a schematic that illustrates the optical spectra of the comb 1 (in blue) and comb 2 (in red). The interference of the two combs creates a series of beat frequencies after proper filter. The lower part of the figure depicts the CLS algorithm schematic. After the interferogram undergoes Fourier transformation and is normalized against the initial interferogram, the resulting pattern theoretically forms a spiral line within the space composed of the complex plane and the frequency axis.}
\label{fig:myfigure}
\end{figure}

Optical two-way time-frequency transfer (O-TWTFT) utilizes linear optical sampling (LOS) to achieve femtosecond-accurate synchronization of remote clocks through the computation of precise time differences. As shown in Fig.~\ref{fig:pic1}, we employ two frequency combs—a local comb operating at a repetition frequency \( f_r \) and an incoming comb with a deviation of \( \Delta f_r \). Heterodyne detection generates time-stamped interferograms at a rate of \( \Delta f_r \), each of which is subjected to a Fourier transformation to produce its spectral phase \( \theta(\nu, n) \). Here, \( \nu \) denotes the Fourier frequency, and \( n \) indicates the \( n \)-th interferogram. The initial interferogram \( \theta(\nu, 0) \) acts as a reference, and the phase difference \( \theta(\nu, n) - \theta(\nu, 0) \) is linearly related to the Fourier frequency \( \nu \), with the slope encoding the target time difference.

The field of one-dimensional phase unwrapping~\cite{kong20221d} presents a research gap. Itoh's algorithm~\cite{itoh1982analysis}, while commonly used, experiences a rapid decline in performance in the presence of system noise. Alternative methods, such as those based on Kalman filtering~\cite{zhang2019robust, kulkarni2020simultaneous} or Deep Learning~\cite{dmitrieva2020short, kong20221d}, are computationally demanding and not optimized for O-TWTFT. Even the algorithm developed by Shen et al.~\cite{shen2021experimental}, which was more suitable, struggles under significant noise.

The Fourier transform translates the temporal information of interferograms into the frequency domain, characterized by both phase and amplitude. Prior phase unwrapping algorithms for O-TWTFT in Shen et al.~\cite{shen2021experimental} focused solely on phase, neglecting amplitude information. We introduce a novel phase unwrapping algorithm that capitalizes on both the imaginary and real parts of the frequency domain, thus utilizing the full informational content of interferograms. It is a well-established fact that white noise retains its characteristics in both real and imaginary components after a discrete Fourier transformation. Moreover, under white noise conditions, the least squares fitting is equivalent to maximum likelihood estimation (MLE)~\cite{charnes1976equivalence}. MLE is prized for its mathematical properties, including consistency, asymptotic normality, and efficiency~\cite{casella2021statistical}. Leveraging these attributes, we apply a least squares fitting method to the complex results of the Fourier-transformed interferograms, which is equivalent to MLE in this context.

As depicted in Fig.~\ref{fig:myfigure}(b), the Fourier-transformed interferogram spectrum is represented in three-dimensional space, with frequency, real component, and imaginary component along the x, y, and z axes, respectively. In a noise-free scenario, the phase difference linearity theoretically results in a helical curve when the interferogram is processed against the initial one. This helical curve is defined by three parameters: the attenuation coefficient \( \beta \), the initial phase difference term \( \gamma \), and the common difference \( \alpha \). The curve fitting using the least squares method enables the calculation of the one-way delay through the slope \( \alpha \) and provides the amplitude coefficient \( \beta \), which is crucial for determining the received power of the interferogram. A detailed explanation and further elaboration of this algorithm are provided in the Supplementary Material. This algorithm not only preserves the complete information of the interferograms but also exploits the favorable properties of MLE, significantly enhancing the precision of the estimated time differences.

\section{EXPERIMENTAL SETUP}

\begin{figure*}[htbp]
\includegraphics[width=17.6cm]{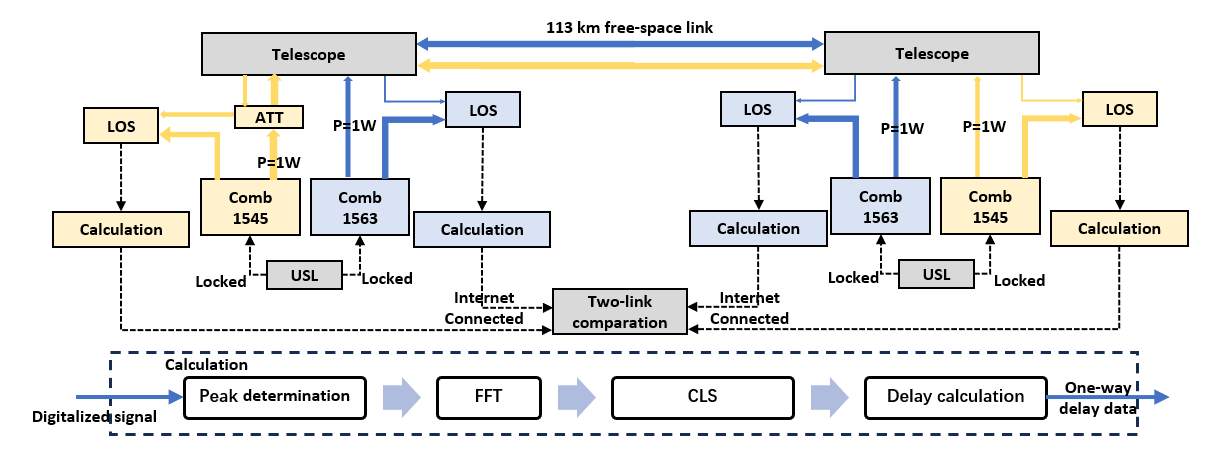}
\caption{Experimental Setup: The schematic illustrates two free-space links, with the 1,545-nm link in yellow and the 1,563-nm link in blue. The lower section details the computational steps within the computation module, including the phase unwrapping process performed using both the CLS algorithm and the previous algorithm~\cite{shen2021experimental} for comparative analysis.}\label{fig:exp}
\end{figure*}

To rigorously evaluate the performance of our proposed CLS algorithm, especially under low-SNR scenarios, we conducted a series of experiments in Urumqi, Xinjiang Province, to assess time-frequency transfer across a high-loss link. We compared the CLS algorithm with the previous algorithm by Shen et al.~\cite{shen2021experimental} using the same dataset of interferogram data from this high-loss link. Our experimental configuration closely aligns with that reported in Shen et al. \cite{shen2022free}. As shown in Fig.~\ref{fig:exp}, the two terminals, situated in Nanshan and Gaoyazi, are 113 km apart and are equipped with an ultra-stable laser (USL), two high-power OFCs, two LOS modules, two computational modules, and optical transceiver telescopes. By employing frequency multiplexing within a shared free-space channel, we established two independent two-way time-frequency transfer links. We intentionally increased the loss on the 1,545-nm link to generate a dataset for algorithm comparison, using the 1,563-nm link with lower loss as a control.

Each terminal houses a USL with a frequency instability of $3\times10^{-15}$ at 1 second and a wavelength ($\lambda$) of 1,550.12 nm, serving as the reference clock source. The OFCs, optically phase-locked to the USL, function as carrier and reference signals. The 1,545-nm OFC operates at a 250 MHz repetition rate with a 2.5 kHz differential repetition rate between terminals, while the 1,563-nm OFC has a 200 MHz repetition rate with a 2 kHz differential rate. To counteract the significant link loss, we utilized a two-stage high-power erbium-doped fiber amplifier (EDFA), achieving a 1-W output power with a 20-nm filtered spectrum. The 3-dB bandwidth of the OFCs at both wavelengths was restricted to approximately 7 nm by the gain region of the EDFA. In the LOS modules, interference data is detected, digitized, and sent to the calculation module for processing. The optical transceiver telescopes, with automatic direction tracking, use an orthogonal polarization scheme to differentiate between sending and receiving OFCs, providing substantial isolation in two-way transfer and mimicking a satellite-to-ground scenario with relative angular motion distinguishing backward and forward transfers.

To emulate ground-satellite link conditions, we implemented two strategies to elevate the 1,545-nm link's loss. First, an attenuator at the Nanshan end of the 1,545-nm link reduced the received power of interferograms at both terminals. Second, intentional misalignment of the optical path further decreased the received power of interferograms on the 1,545-nm link.

The heightened link loss at 1,545 nm led to a reduced SNR, complicating the extraction of valid data. Further details on the data filtering process are provided in the Supplementary Materials.

The 1,563-nm link demonstrates an average link loss of approximately 75 dB. The performance at such loss has been validated in previous experiments~\cite{shen2022free}. In contrast, the 1,545-nm link experiences a significantly higher average link loss, estimated at 90 dB. Link losses are calculated based on the received power of the interferograms, as detailed in the Supplemental Materials. We applied both algorithms to the interferogram dataset from the high-loss 1,545-nm link, while the previous algorithm was utilized for the 1,563-nm link due to its established effectiveness. The differences in outcomes between the two links highlight the inherent errors present in the 1,545-nm link.

\section{Results}

\begin{figure}[htbp]
  \centering
  \sidesubfloat[]{\includegraphics[width=0.9\columnwidth]{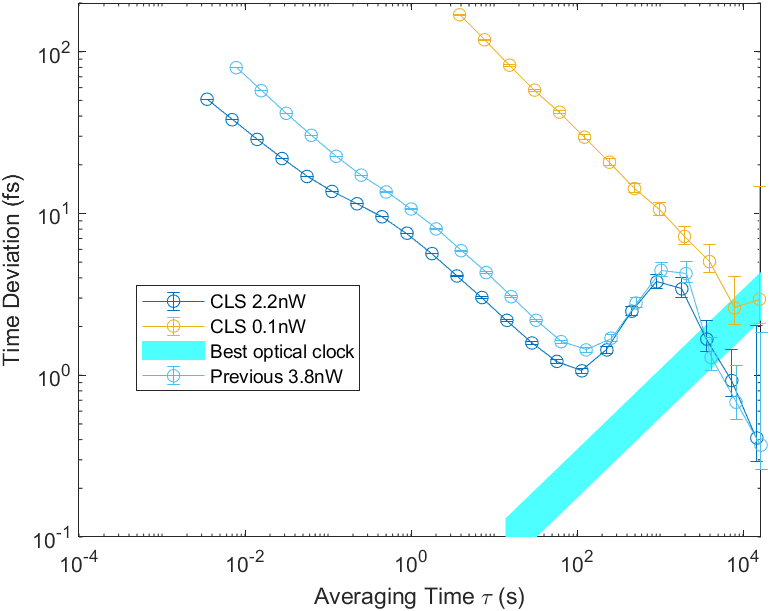}
\label{fig:tdev}}\\
  \sidesubfloat[]{\includegraphics[width=0.9\columnwidth]{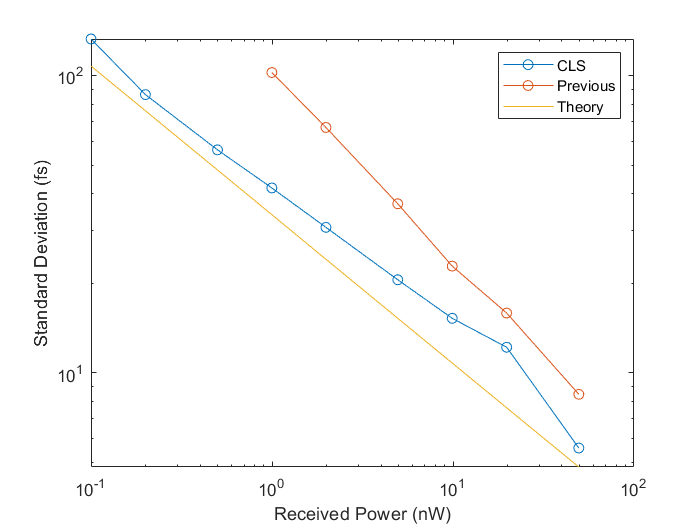}}\label{fig:std}
\caption{(a) Time deviation (TDEV) curves for the CLS and previous algorithms. The CLS algorithm outperforms the previous algorithm by half an order of magnitude. The average received power for the previous algorithm is 3.8 nW, while for the CLS algorithm, it is 2.2 nW, reflecting their respective minimum received power capabilities. The TDEV curve for the CLS algorithm at 0.1-nW filtered data parallels the other curves over long integration times, indicating negligible long-term stability impact. Performances of top optical clocks are also shown for comparison~\cite{oelker2019demonstration, schioppo2017ultrastable}. (b) The relationship between standard deviation and received power is depicted, with the blue line for the CLS algorithm and the orange line for the previous algorithm. The CLS algorithm's performance approaches the quantum noise limit, represented by the yellow line.}\label{fig:result}
\end{figure}

Fig.~\ref{fig:result}(a) presents the time deviation (TDEV) for both the CLS and previous algorithms. Over short averaging intervals, the TDEV of both algorithms exhibits a downward trend, with the slope being proportional to the square root of the averaging time ($\tau$). Notably, the CLS algorithm demonstrates greater stability than the previous algorithm at each averaging point. The CLS algorithm's TDEV starts at 50 fs and improves to 1 fs within approximately 100 seconds, whereas the previous algorithm begins at a higher TDEV of 80 fs. The CLS algorithm also collects more valid data, with an average data rate of 280 samples per second, compared to 130 samples per second for the previous algorithm, which struggles with data having a received power less than 1 nW. For averaging periods less than 100 seconds, the previous algorithm fails to match the performance of the CLS algorithm. Moreover, when benchmarked against the best atomic clocks, the CLS algorithm-linked performance shows superior long-term stability. To explore the performance limits of the CLS algorithm, we also present the TDEV curve at 0.1-nW power, indicating that even at low data rates, the CLS algorithm's TDEV can achieve sub-femtoseconds at 10,000 seconds of averaging. Details on selecting specific received power and filtering valid data can be found in the Supplemental Materials.

Fig.~\ref{fig:result}(b) illustrates the standard deviation of the results for both algorithms. As signal power decreases, the SNR of the interferograms diminishes, leading to an increase in the standard deviation of the results. The CLS algorithm's curve is consistently lower than that of the previous algorithm, indicating its effectiveness in reducing estimation errors under low SNR conditions. In extremely low SNR scenarios, both algorithms may produce erroneous results that deviate significantly from the true value. We define the minimum received power for an algorithm as the threshold at which 99\% of its output data falls within three standard deviations. The CLS algorithm's minimum received power is approximately 0.1 nW, while the previous algorithm's is about 1 nW. Figure~\ref{fig:result}(b) also displays the fundamental quantum limit, given by the equation~\cite{ellis2021scaling, richards2005fundamentals}:

\begin{equation}
\sigma_t^2 \approx \frac{\tau_p^2}{\text{SNR}}
\label{eq:limit}
\end{equation}

where \(\tau_p\) represents the intensity full width half maximum (FWHM) of the unchirped frequency comb pulses, and SNR is the signal-to-noise ratio of the receiver. Under low received power conditions, instabilities approach the quantum noise limit defined by the equation.

\section{conclusion}
The CLS marks a significant advancement in phase-unwrapping methods, particularly excelling under conditions of low received power and SNR. This algorithm not only reduces fluctuations but also enhances precision, substantially improving the robustness and accuracy of time-frequency transfer in optical two-way time-frequency transfer (O-TWTFT) systems. By employing the CLS algorithm, the data rate is augmented from 130 to 280 samples per second, and the standard deviation is reduced from 80 femtoseconds to 50 femtoseconds. The minimum received power requirement is also significantly lowered from 1~nW to 0.1~nW. This enhanced detection sensitivity extends the operational range of the LOS system by more than threefold. With a Watt-level OFC, our CLS algorithm can endure losses exceeding 100 dB, making it viable for Earth-moon link applications. In Earth-orbit satellite-to-ground scenarios, this translates to the potential use of less powerful OFCs and smaller telescopes, thereby optimizing resource efficiency. Notably, our method requires no hardware modifications, representing a pivotal step forward in the field of ground-to-satellite time-frequency dissemination.

\vspace{\baselineskip}

This research was supported by the National Key Research and Development Programme of China (grant no. 2020YFC2200103, 2020YFA0309800); National Natural Science Foundation of China (grant no. T2125010); Strategic Priority Research Programme of Chinese Academy of Sciences (grant no. XDB35030000); Anhui Initiative in Quantum Information Technologies (grant no. AHY010100); Key R\&D Plan of Shandong Province (grant no. 2021ZDPT01); Shanghai Municipal Science and Technology Major Project (grant 2019SHZDZX01); Innovation Programme for Quantum Science and Technology (grant no. 2021ZD0300100, 2021ZD0300300, 2021ZD0300903).

Y.-C. Fang and J.-Y. Guan contributed equally to this work.


%
  
\begin{titlepage}

\centering
{\LARGE\bfseries Supplemental Material}
\vspace{2cm}
\end{titlepage}
\preprint{APS/123-QED}

\title{Supplemental Material: Enhanced timing of a 113 km O-TWTFT link with digital maximum likelihood estimation process}
\author{Yu-Chen Fang}
\affiliation{Hefei National Research Center for Physical Sciences at the Microscale and School of Physical Sciences, University of Science and Technology of China, Hefei, China}
\affiliation{Shanghai Research Center for Quantum Sciences and CAS Center for Excellence in Quantum Information and Quantum Physics, University of Science and Technology of China, Shanghai, China}
\affiliation{Hefei National Laboratory, University of Science and Technology of China, Hefei, China}
\author{Jian-Yu Guan}
\affiliation{Shanghai Research Center for Quantum Sciences and CAS Center for Excellence in Quantum Information and Quantum Physics, University of Science and Technology of China, Shanghai, China}
\affiliation{Hefei National Laboratory, University of Science and Technology of China, Hefei, China}
\author{Qi Shen}
\author{Jin-Jian Han}
\affiliation{Hefei National Research Center for Physical Sciences at the Microscale and School of Physical Sciences, University of Science and Technology of China, Hefei, China}
\affiliation{Shanghai Research Center for Quantum Sciences and CAS Center for Excellence in Quantum Information and Quantum Physics, University of Science and Technology of China, Shanghai, China}
\affiliation{Hefei National Laboratory, University of Science and Technology of China, Hefei, China}
\author{Lei Hou}
\affiliation{Shanghai Research Center for Quantum Sciences and CAS Center for Excellence in Quantum Information and Quantum Physics, University of Science and Technology of China, Shanghai, China}
\affiliation{Hefei National Laboratory, University of Science and Technology of China, Hefei, China}
\author{Meng-Zhe Lian}
\affiliation{Hefei National Research Center for Physical Sciences at the Microscale and School of Physical Sciences, University of Science and Technology of China, Hefei, China}
\affiliation{Shanghai Research Center for Quantum Sciences and CAS Center for Excellence in Quantum Information and Quantum Physics, University of Science and Technology of China, Shanghai, China}
\affiliation{Hefei National Laboratory, University of Science and Technology of China, Hefei, China}
\author{Yong Wang}
\affiliation{Xinjiang Astronomical Observatory, Chinese Academy of Sciences, Urumqi 830011, China}
\author{Wei-Yue Liu}
\affiliation{Shanghai Research Center for Quantum Sciences and CAS Center for Excellence in Quantum Information and Quantum Physics, University of Science and Technology of China, Shanghai, China}
\affiliation{Hefei National Laboratory, University of Science and Technology of China, Hefei, China}
\affiliation{Faculty of Information Science and Engineering, Ningbo University, Ningbo 315211, China}
\author{Ji-Gang Ren}
\author{Cheng-Zhi Peng}
\author{Qiang Zhang}
\author{Hai-Feng Jiang}
\author{Jian-Wei Pan}
\affiliation{Hefei National Research Center for Physical Sciences at the Microscale and School of Physical Sciences, University of Science and Technology of China, Hefei, China}
\affiliation{Shanghai Research Center for Quantum Sciences and CAS Center for Excellence in Quantum Information and Quantum Physics, University of Science and Technology of China, Shanghai, China}
\affiliation{Hefei National Laboratory, University of Science and Technology of China, Hefei, China}

\maketitle
\section{\label{sec:level1}CLS Algorithm}

\subsection{Comprehensive Description of the CLS Algorithm}
\begin{figure}[ht]
    \centering
    \includegraphics[width=8.6cm]{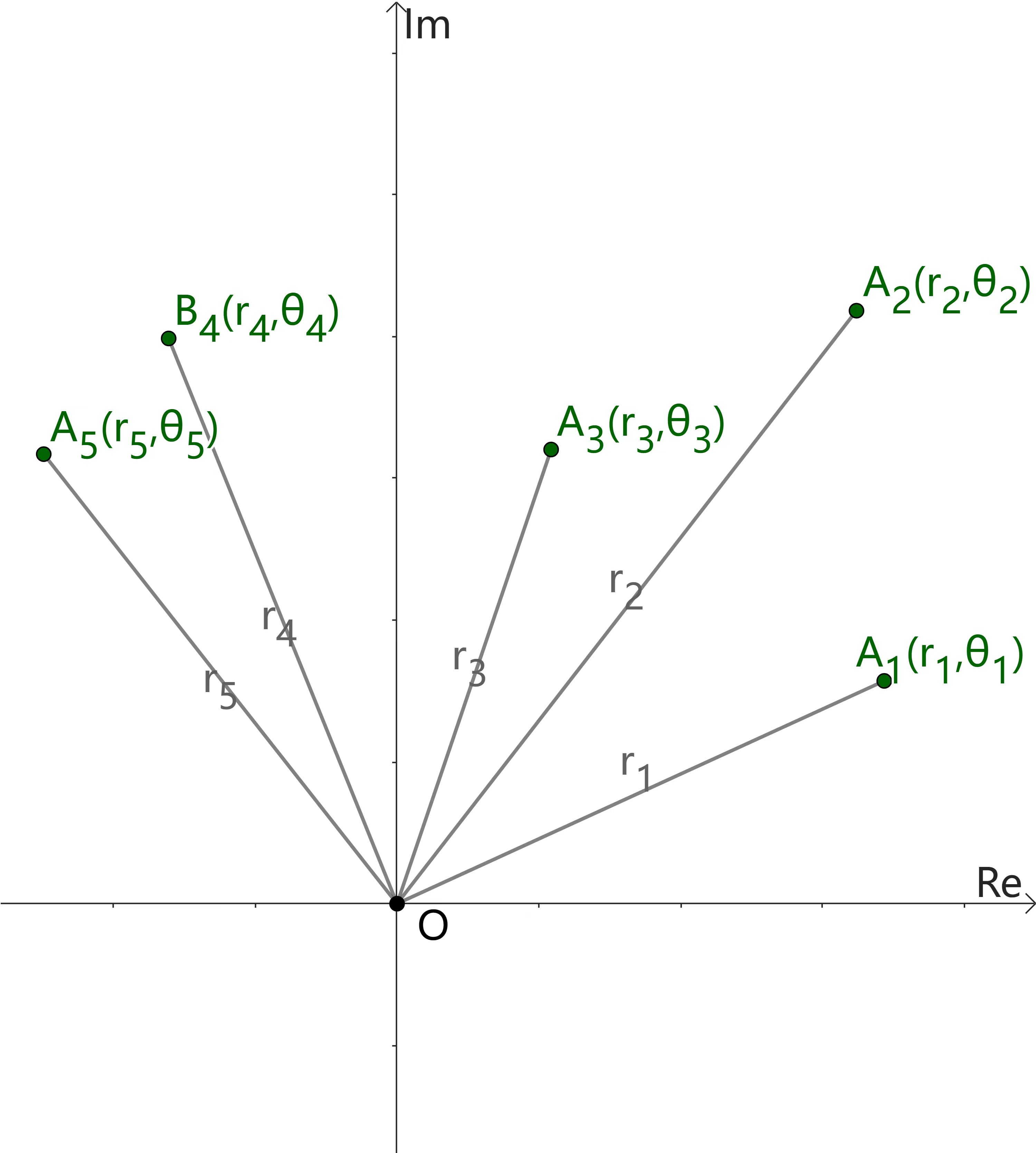}
    \caption{Depiction of initial frame data points on the complex plane post-Fourier transformation. Green points $A_1, A_2, A_3, A_4, A_5, \ldots$ are presented with their respective magnitudes $r_1, r_2, r_3, r_4, r_5, \ldots$ and phase angles $\theta_1, \theta_2, \theta_3, \theta_4, \theta_5, \ldots$.}
    \label{fig:initial_frame}
\end{figure}

\begin{figure}[ht]
    \centering
    \includegraphics[width=8.6cm]{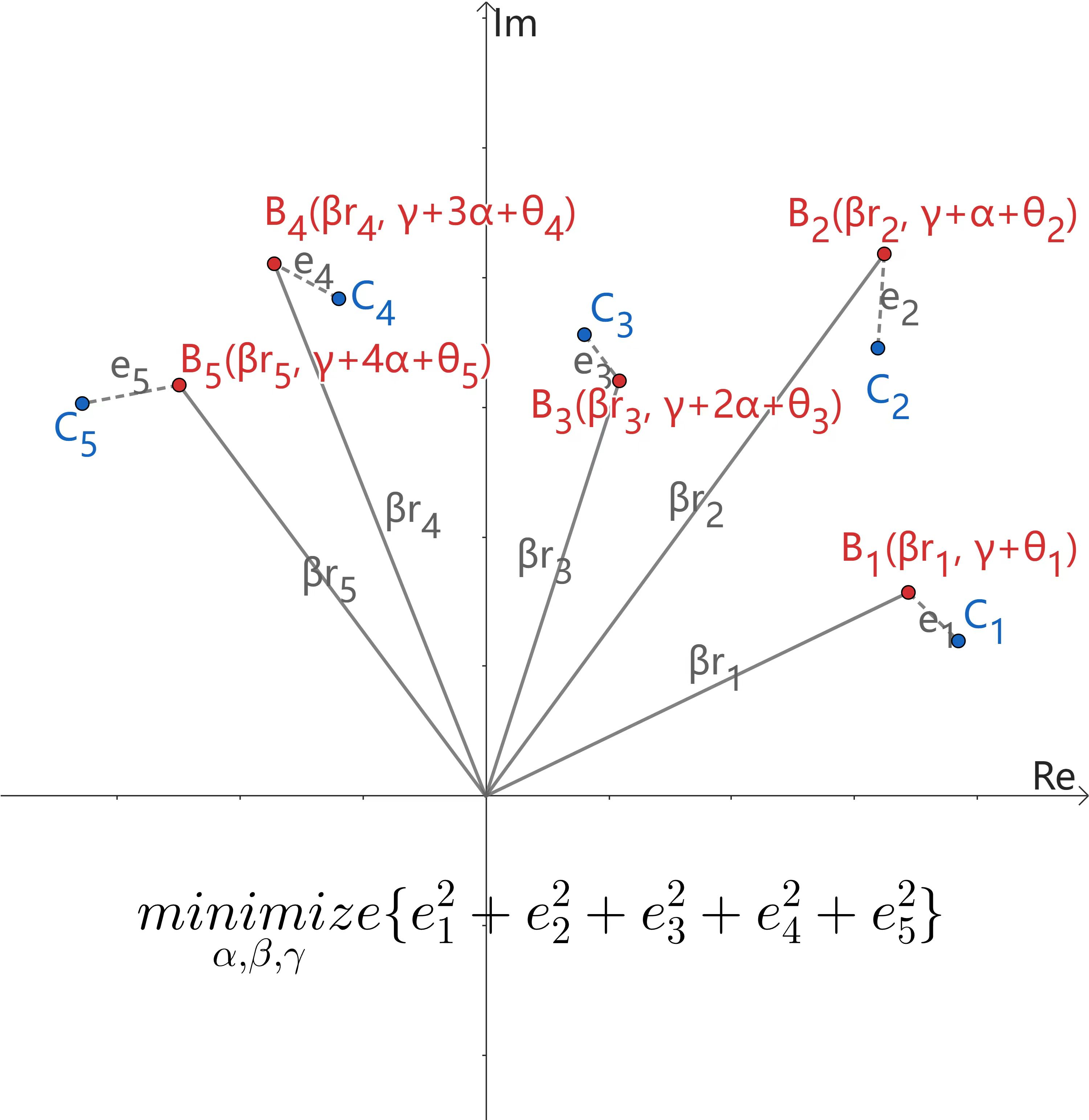}
    \caption{Illustration of the CLS algorithm in action, applied to discrete Fourier transform data. Predicted values are denoted by red points \( B_k \), and actual data points are shown in blue \( C_k \), with the residuals $e_k$ indicating the discrepancy. The algorithm's objective is to minimize the sum of the squares of these residuals, represented here in the complex plane, to refine the signal's parameter estimation.}
    \label{fig:CLS_algorithm}
\end{figure}

To compute the delay of current interferograms, our approach involves constructing a model with delay as a parameter and then minimizing the discrepancy between the model's predicted values and the sampled interferograms in the frequency domain. This method facilitates an accurate estimation of the delay. To build a predictive model for the interferograms, we require the amplitude and phase information from the initial interferogram. As illustrated in Fig.~\ref{fig:initial_frame}, the initial frame is mapped onto the complex plane following a Fourier transformation. This yields a series of points denoted as \( A_1, A_2, A_3, A_4, A_5, \ldots \), each characterized by their respective amplitudes \( r_1, r_2, r_3, r_4, r_5, \ldots \) and phase angles \( \theta_1, \theta_2, \theta_3, \theta_4, \theta_5, \ldots \). Fig.~\ref{fig:CLS_algorithm} describes the parameters and fitting process of the CLS algorithm. We consider each point obtained after the discrete Fourier transformation as a sample, and the points \( C_1, C_2, C_3, C_4, C_5, \ldots \) represent our sampling results. We now aim to estimate the parameters of our model from these sampling results. Our model has three parameters, which physically signify the changes in amplitude and phase of the current interferogram relative to the initial interferogram. Specifically, \( \beta \) denotes the coefficient of amplitude attenuation, while \( \alpha \) and \( \gamma \) characterize the phase changes. According to the theory of linear optical sampling\cite{giorgetta2013optical}, the phase difference between the current frame and the initial frame forms an arithmetic sequence to frequency; hence only two parameters are needed: \( \alpha \) represents the common difference of the arithmetic sequence, and \( \gamma \) represents the initial phase, completely characterizing the phase change. Once the parameters of the model are established, we can proceed to predict values for each sample, which are represented by the points \( B_1, B_2, B_3, B_4, B_5, \ldots \). For instance, the amplitude of \( B_3 \) is given by \( \beta r_3 \), and its phase angle by \( \gamma + 2\alpha + \theta_3 \). The differences between the predicted values and the sampled values are denoted as \( e_k \). Our objective is to ensure that the predicted points are as proximate to the sampling points as possible. To achieve this, we apply the least squares method, which aims to minimize the sum of the squares of the residuals, expressed by the following equation:
\begin{equation}
\mathop{\mathrm{minimize}}\limits_{\alpha, \beta, \gamma} \{\sum_k e_k^2\}
\end{equation}

Herein, we describe the specific steps of the CLS algorithm:
 
\begin{enumerate}
    \item \textbf{Model Establishment}:\\
    Perform Discrete Fourier Transforms on both the initial and current frames to map the amplitude and phase information onto the complex plane. A model is then established to encapsulate the amplitude and phase variations of the current interference frame with the initial frame, parameterized by \( \alpha \), \( \beta \), and \( \gamma \), which denote the phase change rate, amplitude attenuation coefficient, and initial phase offset, respectively.
    
    \item \textbf{Constructing the Objective Function}:\\
    Formulate the objective function as the sum of the squares of the residuals between the model's predicted values and the observed current frame values:
    \begin{equation}
        L(\alpha, \beta, \gamma) = \sum_{k} \left| B_k - C_k \right|^2
    \end{equation}
    where \( B_k \) represents the model's predicted complex values of the $k$-th point, and \( C_k \) denotes the $k$-th point of Discrete Fourier Transform results of the current frame.
    
    \item \textbf{Minimizing the Objective Function}:\\
    Implement an iterative minimization of the objective function using Newton's method. To mitigate the risk of convergence to local minima, multiple initial starting points are used for the iterations.
    
    \item \textbf{Outcome}:\\
    After the completion of the fitting process, use the determined parameter \( \alpha \) to calculate the time delay and parameter \( \beta \) to compute the amplitude or received power of the interferograms. The initial phase \( \gamma \) is also derived, which corresponds to the carrier phase of the current frame.
\end{enumerate}

\subsection{Theoretical Analysis of the CLS Algorithm}
\label{sec:cls_theory}
In this section, we theoretically discuss the properties of the CLS algorithm, demonstrating that it is, in fact, a Maximum Likelihood Estimation (MLE) of the parameters.

A fundamental assumption in our analysis of our model is that the noises present in the data can be modeled as Gaussian noise.  

According to the literature on the discrete Fourier transform of noise \cite{richards2013discrete}, a time-domain white noise transforms into white noise along the real and imaginary axes on the complex plane after undergoing a discrete Fourier transform. Hence, when dealing with complex data, the noise to consider remains Gaussian white noise.

Furthermore, as per literature that equates least squares estimation with MLE under the premise of independently distributed Gaussian white noise \cite{charnes1976equivalence}, the two estimations are equivalent.

Under certain regularity conditions, theoretical statistical inference posits that MLE yields asymptotic normal, consistent and asymptotic efficient results \cite{casella2021statistical}. 
\begin{enumerate}
  \item \textbf{Asymptotic Normality:} This concept states that with increasing sample size, the distribution of the MLE will approximate a normal distribution.
  
  \item \textbf{Asymptotic Consistency:} Asymptotic Consistency implies that the expected value of the MLE converges to the true parameter value as the sample size increases.
  
  \item \textbf{Asymptotic Efficiency:} Asymptotic efficiency denotes that the MLE, among all unbiased estimators, will have the smallest possible variance when the sample size is large.
\end{enumerate}

We have previously established that the CLS method is equivalent to the MLE, and it can be verified that our estimation concerning the delay time adheres to these regularity conditions. On another note, our experimental setup is capable of generating hundreds or even thousands of samples per second, easily fulfilling the sample size requirement for the asymptotic properties.

In summary, the CLS algorithm proposed for computing delay time demonstrates the following advantageous statistical characteristics:
\begin{enumerate}
    \item The results exhibit a normal distribution, indicating that outliers are uncommon.
    \item Bias is absent in the estimation of delay time, ensuring the reliability of the estimation.
    \item Among all possible unbiased algorithms, CLS achieves the lowest error variance, signifying its superior precision.
\end{enumerate}

.

\section{Data Filtering}

In our experiment, due to the introduction of attenuators, the link loss for the 1,545-nm link was increased, potentially obscuring the interference patterns generated by the optical comb with environmental noise. In such circumstances, it posed a significant challenge for the FPGA to discern whether the received signals were pure noise or contained valid interference information. Hence, in our experimental design, we decided not to rely on FPGA to make judgments on signal validity but instead to collect all signals received by FPGA in their entirety. As a result, our collected dataset included a substantial amount of noise components, necessitating meticulous data filtering and analysis in subsequent data processing steps.

The steps for filtering valid frames are divided into three stages, which will be discussed in detail later, demonstrating the feasibility and effectiveness of these steps and methods.

\begin{enumerate}
\item Select high-SNR interferograms (signal power above 5~nW) and calculate the delay.

\item Estimate the one-way delay of each sampling time by interpolating the high-SNR delay data.

\item Calculate the one-way delay for each interferogram; if it deviates by more than 1000~fs from the delay estimated from the sampling timestamp of that interferogram, it is deemed a noise frame.
\end{enumerate}

\begin{figure}[h]
    \centering
    \includegraphics[width=8.6cm]{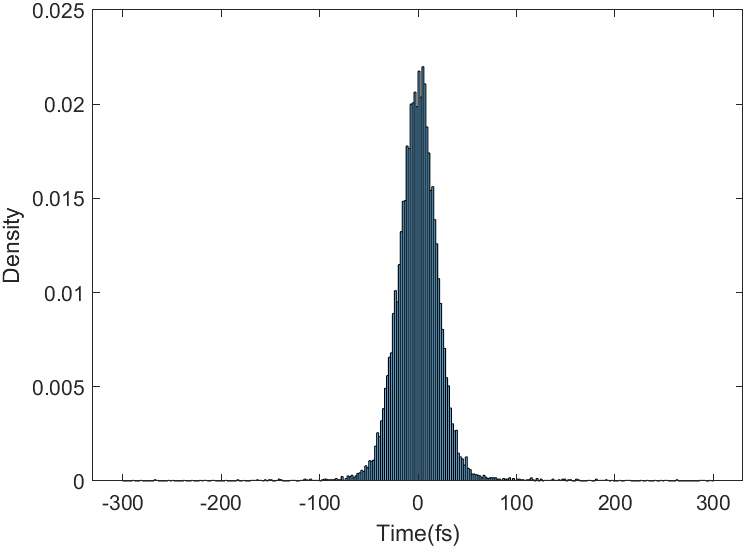} 
    \caption{The distribution of time differences between the actual one-way delays calculated by interferograms and the estimated one-way delays obtained using linear interpolation of timestamps, which follows a normal distribution.}
    \label{fig:normal_distribution}
\end{figure}

First, we justify the feasibility of selecting high SNR interferograms. According to the findings in \cite{shen2022free}, the received power is log-normally distributed due to atmospheric fluctuations. That is to say, despite the overall significant link loss, it is still possible to select some interferograms with relatively high SNR. Fig.\ref{fig:normal_distribution} illustrates the distribution of received power from the sampling data, showing that we can choose interferograms with a received power above 5~nW. In this experiment, an average of more than 4 high-SNR interferograms per second could be selected, and the delays calculated from these frames were relatively accurate, facilitating subsequent calculations and estimations.

\begin{figure}[h]
    \centering
    \includegraphics[width=8.6cm]{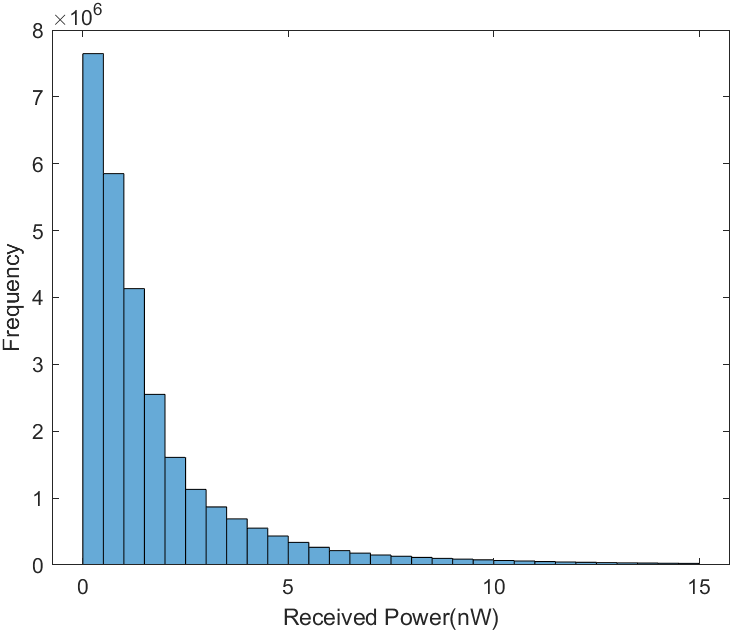} 
    \caption{The distribution of received optical power.}
    \label{fig:received_power}
\end{figure}

After calculating the one-way delay for high SNR interferograms, we can estimate the approximate value of the one-way delay for each timestamp. The relationship between the calculated time points and one-way delay, as shown in Fig.~\ref{fig:received_power}, suggests that we can use linear interpolation to estimate the delay corresponding to the timestamps. The linearity of the one-way delay is due to the fixed offset between the used ultra-stable laser frequency and the real frequency; the fixed frequency offset manifests as a linear drift in delay. In the calculation of clock differences, the discrepancies at both ends will be completely canceled out.

Following the estimation of delays, we further compare them with the calculated delays. If the difference between the two exceeds 1000~fs, the data received will be considered a noise frame by a false trigger, not containing any valid signal components. We have rigorously demonstrated at a theoretical level that under our experimental conditions, the difference between the estimated delay and the actual delay will almost certainly be less than 1000~fs.

In an experimental dataset collected without the inclusion of an attenuator, ensuring that the received light intensity of even the weakest signals exceeds 5~nW, we posit that the delay data are fundamentally accurate. Utilizing the same methodology, four data points are collected per second, and each timestamp's delay is estimated through linear interpolation. As shown in Fig.~\ref{fig:normal_distribution}, the time differences, calculated by subtracting the estimated delay from the actual interferogram delay, approximately follow a normal distribution with a standard deviation of about 40~fs.

This result underscores that the standard deviation of the discrepancy in our estimation of the one-way delay using high-intensity data is about 40~fs. Given this standard deviation, the probability of the true delay deviating from the estimated delay by more than 1000~fs is exceedingly low.

Theoretical analysis also scrutinizes the fluctuations in delay. The actual one-way delay is predominantly influenced by two factors: the frequency drift of the ultra-stable laser and the pulse's flight time. For this study, with a high SNR data acquisition rate of about 4 data points per second, it suffices to demonstrate that within such brief temporal intervals, the delay variations induced by these factors are exceedingly unlikely to surpass 1000~fs.

Firstly, according to experimental measurements, the drift rate of the ultra-stable laser frequency is less than 1~Hz per second. Therefore, within a 0.25-second window, considering the operating frequency of the ultra-stable laser is about 200~THz, the time delay impact due to the laser's frequency drift will be less than 5~fs.

Secondly, the fluctuation of the pulse flight time can be calculated using an atmospheric model. Based on the atmospheric model obtained from our hundred-kilometer experiment~\cite{shen2022free}, the power density of the atmosphere approximates to \(486.4f^{-\frac{8}{3}}\) over a large range.

Knowing from the total data that the amount of high SNR data is about 4 per second, here we conservatively calculate atmospheric fluctuations above 3~Hz. From the properties of the power spectral density, we know that the delay fluctuations caused by the atmosphere greater than 3~Hz can be described as follows:
\[
    RMS = \sqrt{\int_{3}^{\infty} 486.4f^{-\frac{8}{3}} \, df}
\]
The calculation yields an RMS of 46~fs for atmospheric fluctuations within this frequency range. 

Previously, we mentioned that the time differences overall exhibit a normal distribution. Given that the impact of frequency drift can be neglected over short periods, this indicates that the remaining fluctuations due to atmospheric turbulence follow a normal distribution. The probability of a delay error exceeding 155 fs is 1/1000. A delay fluctuation of 1000 fs is approximately 21 standard deviations, making it an almost impossible event for a reasonable result.

Taking both factors into account, it is clear that within the 0.25-second interval, the delay variations caused by both the laser frequency drift and the atmospheric turbulence are significantly lower than 1000fs. Thus, we confidently filter out noise frames using the threshold of 1000fs, providing robust data for the subsequent calculation of the time difference between the two sites.

Our filtering process has proven effective in the face of high noise and loss conditions, ensuring the reliability of time synchronization across optical links. The methods developed here hold promise for optical communication technologies in similarly challenging environments.

\section{Received power calculation}
In this experiment, the ratio between transmitted and received OFC power is up to a few $10^9$, three orders of magnitude higher than the isolation between the sending and receiving optical paths (approximately 60~dB). Consequently, the received power cannot be measured simultaneously while the link is operational. Therefore, we calculate the received optical power using the amplitude coefficient $\beta$ from the CLS algorithm. In the following part, we will introduce this method and demonstrate its effectiveness.

In the laboratory, we used the same setup as in the experiment conducted in Urumqi to obtain a set of interferograms and measured the corresponding received optical power using a commercial power meter. 

\begin{figure}[h]
    \centering
    \includegraphics[width=8.6cm]{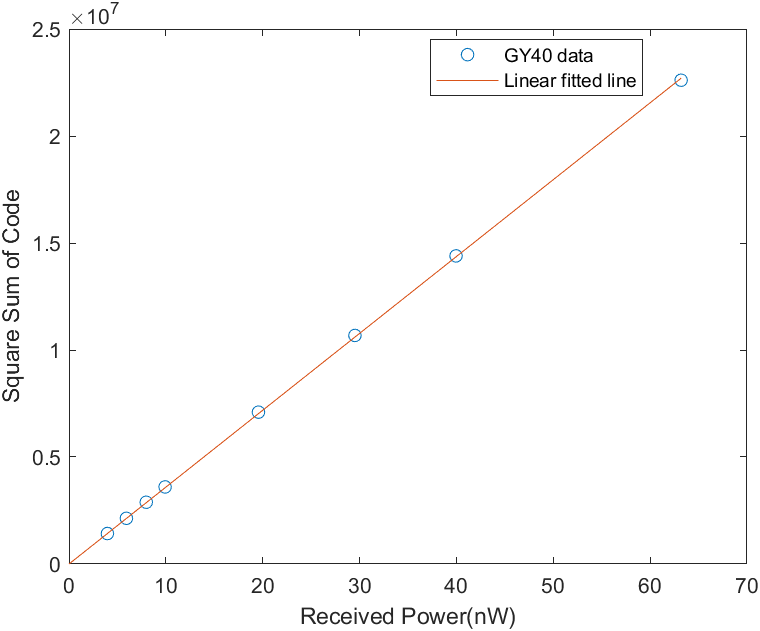} 
    \caption{The relationship between the energy of the interferogram and the received optical power, showing a linear relationship with a non-zero intercept due to the constant power of the Gaussian noise.}
    \label{fig:energy_vs_power}
\end{figure}

We first established the proportionality between the energy of the interferogram and the received optical power. Fig.~\ref{fig:energy_vs_power} illustrates the relationship between the energy and the received optical power, obtained by calculating the sum of the squares of the signal samples. The plot shows a linear relationship with a non-zero intercept.

Noting that the noise in the interferograms is Gaussian with a constant standard deviation, independent of the received optical power, we can straightforwardly deduce that the intercept part is the sum of the squares of the noise. Furthermore, the part proportional to the optical power is the energy of the noise-free signal. 

Therefore, we can leverage this proportionality to determine the received optical power corresponding to the interferograms in the experiment conducted in Urumqi. However, in cases of extremely low SNR, the energy calculated from the sum of the squares of the signal fluctuates significantly. Thus, we first use this method to obtain the received optical power $P_{\mathrm{init}}$ of an initial frame with a relatively high SNR. Then, we use the amplitude coefficient $\beta$ from the CLS algorithm to calculate the received optical power $P_{\mathrm{rec}}$ of the current signal, using the formula $P_{\mathrm{rec}} = P_{\mathrm{init}} \cdot \beta^2$.

\begin{figure}[h]
    \centering
    \includegraphics[width=8.6cm]{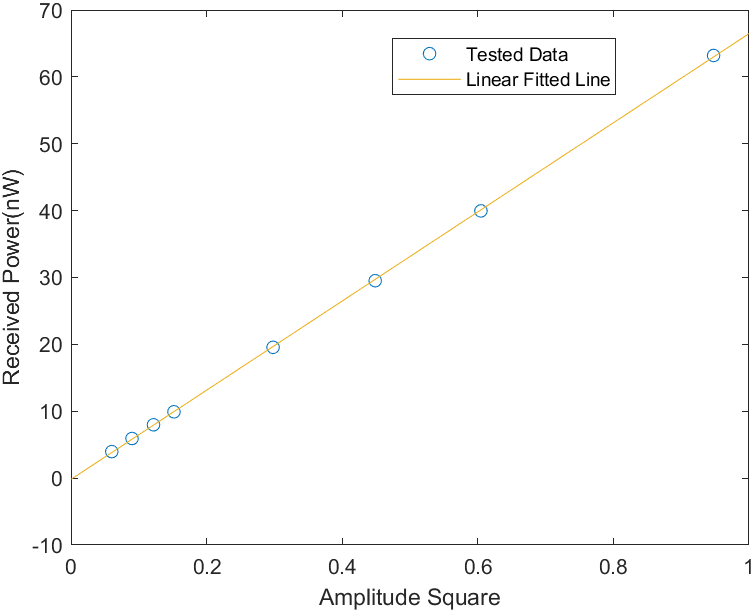} 
    \caption{Depiction of the relationship between the received optical power and the square of the amplitude coefficient, illustrating that the received optical power of an interferogram is proportional to the square of the amplitude coefficient.}
    \label{fig:amplitude_vs_measured_power}
\end{figure}

To verify the effectiveness of using the amplitude coefficient to calculate the received optical power, we applied this method to the laboratory data that included the received optical power. The results, shown in Fig.~\ref{fig:amplitude_vs_measured_power}, indicate that the calculated optical power matches the experimentally measured power very well.

\end{document}